\documentstyle[12pt]{article}
\begin{document}
\def\om{\omega}
\def\omt{\tilde{\omega}}
\def\ti{\tilde}
\def\o{\Omega}
\def\bchi{\bar\chi^i}
\def\In{{\rm Int}}
\def\ba{\bar a}
\def\w{\wedge}
\def\ep{\epsilon}
\def\k{\kappa}
\def\Tr{{\rm Tr}}
 \def\btd{\bigtriangledown}
\def\ST{{\rm STr}}
 \def\hr{ \bigg{\vert}^{\pi}_0}
\def\ss{\subset}
\def\ot{\otimes}
\def\bc{{\bf C}}
 \def\bul{{1\over 2\pi}\int_{-\pi}^{\pi}d\si}
  \def\bdr{{1\over \pi}\int_{0}^{\pi}d\si}
\def\br{{\bf R}}
\def\de{\delta}
\def\tr{\triangleleft}
\def\al{\alpha}
\def\la{\langle}
 \def\omm{\om_{m_\pi m_0}}
\def\ra{\rangle}
\def\G{\Gamma}
\def\st{\stackrel{\wedge}{,}}
\def\sto{\stackrel{\otimes}{,}}
\def\th{\theta}
\def\lm{\lambda}
\def\U{\Upsilon}
\def\jp{{1\over 2}}
\def\js{{1\over 4}}
\def\d{\partial}
\def\tr{\triangleright}
\def\trl{\triangleleft}

\def\be{\begin{equation}}
\def\ee{\end{equation}}
\def\bea{\begin{eqnarray}}
\def\eea{\end{eqnarray}}
\def\D{{\cal D}}
\def\E{{\cal E}}
\def\G{{\cal G}}
\def\H{{\cal H}}
\def\R{{\cal R}}
\def\T{{\cal T}}
\def\bT{\bar{\cal T}}
\def\F{{\cal F}}
\def\n{{1\over n}}
\def\si{\sigma}
\def\ta{\tau}
\def\ov{\over}
\def\l{\lambda}
\def\L{\Lambda}

\def\pih{\hat{\pi}}
\def\Vt{V^{\ti}}
\def\Ut{U^{\ti}}
\def\e{\varepsilon}
\def\b{\beta}
\def\ga{\gamma}

\begin{titlepage}
\begin{flushright}
{}~
IML 2005-09\\
hep-th/0505240
\end{flushright}

\vspace{3cm}
\begin{center}
{\Large \bf  Nested T-duality}\\ 
[50pt]{\small
{\bf C. Klim\v{c}\'{\i}k }
\\ ~~\\Institute de math\'ematiques de Luminy,
 \\163, Avenue de Luminy, 13288 Marseille, France}

\vspace{1cm}
\begin{abstract}
We identify the obstructions for T-dualizing the boundary WZW 
model and make explicit how they depend on the geometry of  
 branes. In particular, the obstructions disappear for certain  
 brane configurations associated to non-regular elements of 
 the Cartan torus.  It is shown in this case that the boundary  
 WZW model  is "nested" in the twisted boundary WZW model
  as the dynamical subsystem of the latter. 
 \end{abstract}
\end{center}
\end{titlepage}
\newpage

\section{Introduction}

Some time ago, Kiritsis and Obers have proposed a T-dualization
 of the closed string WZW model,
based on   the existence of an outer automorphism of the group
 target $G$.
The purpose of this article is to work out a qualitative remark in
 \cite{KS} where it was suggested that there should exist an  open 
 string version of the Kiritsis-Obers T-duality  \cite{KO}.  In order 
 words, the ordinary and the twisted boundary WZW models 
 should be isomorphic as dynamical systems.  We remind that the former (latter) describes the dynamics of WZW open strings whose end-points stick on ordinary (twisted)  conjugacy classes therefore the existence of such T-duality would  establish quite a non-trivial relation
between branes with different dimensions and geometries.  For generic branes, it turns out, however, that   the duality can be established only at the  price of   constraining some zero modes on each  side of the dual pair of models.  This is the usual state of matters known from
the non-Abelian T-duality  story for  closed strings \cite{KS1,KP} and it therefore seems that, generically, the open strings do not perform better  as their closed counterparts.   Nevertheless, there is a subtle difference
between the open and closed cases hidden in the term "generic". In fact, there are exceptional branes for which the duality can be established at least in one direction, i.e. there is no need to constrain the zero modes  on  one side of the T-dual pair of models. This is a new phenomenon
which never occurs for closed strings and we refer to it as  to the nested or one-way duality,  since, in this case, the full-fledged  non-constrained boundary model  is the physical subsystem of its T-dual.   We shall see that this phenomenon occurs for certain branes associated with the non-regular elements
of the Cartan torus and it is related to the fact that  the centralizers of these elements are sufficiently big.

The plan of the paper is as follows:    first we rewrite  the known  symplectic structure of the (twisted) boundary WZW model     in new variables which single out the zero modes causing the obstruction to duality.  Then  we  show that  in some cases  a small gauge symmetry, coming from the bigger centralizers of non-regular elements, happens  to  gauge away the obstructional zero modes and it  embeds the boundary WZW model into its twisted  counterpart.

\section{Review of the bulk WZW model} 
Usually, the symplectic structure of the  bulk  and also of the boundary WZW model is derived from the second-order least action principle  as in \cite{G, GTT, Gaw}. We shall not repeat this derivation here for the sake of conciseness and we shall rather stick on the Hamiltonian formalism from the very beginning.   Let  $LG$ be the loop group of a Lie group $G$. 
The bulk WZW model is a dynamical system whose phase space   points are  pairs $(g,J)$, $g\in LG$, $J\in Lie(LG)$ and the  symplectic form of which reads  (see \cite{M,K}):
 $$\om=d(J, dgg^{-1})_{\G}+\jp k(dgg^{-1}\st \d_{\si}(dgg^{-1}))_{\G}.\eqno(1)$$
 Here  $dgg^{-1}$ is  a $Lie(G)$-valued  right-invariant  Maurer-Cartan form on $LG$, $\d_{\si}$ is the derivative with respect to the loop parameter $\si$ and
 $$(\chi_1,\chi_2)_{\G}\equiv {1\over 2\pi}\int_{-\pi}^{\pi} d\si \Tr (\chi_1(\si)\chi_2(\si)).$$
The dynamics of the bulk WZW model is then given by a (Hamiltonian)  
vector field $v$ defining the time evolution. Explicitely,
$$v=\int^{\pi}_{-\pi} \d_\si  J(\si){\delta\over \delta J(\si)} +{1\over k}\bigtriangledown^L_{J_L}-
{1\over k}\btd^R_{J_R},\eqno(2)$$
where 
$$J_L\equiv J,\quad J_R= -g^{-1}Jg+kg^{-1}\d_\si g.$$
Remind that, for $\chi\in Lie(LG)$, the differential operators $\bigtriangledown^L_\chi$, $\btd^R_\psi$ acts on functions on the loop group $LG$ as follows
$$\btd^L_\chi f(g)={d\over dt} f(e^\chi g)\vert_{t=0}, \quad \btd^R_\psi f(g)={d\over dt} f(ge^\chi)\vert_{t=0}. $$
For completeness, the vector field $v$  fulfills  the relation $\iota_v\om \equiv\om(v,.)=dH$ where the hamiltonian $H$ is given by  (cf.\cite{K})
$$H=- {1\over 4\pi k} \int_{-\pi}^{\pi} d\si \Tr (J_L^2+J_R^2).$$
Finally, let us note that the observables $J_L$ and $J_R$ generate (via the Poisson bracket) the left
and right action of the loop group $LG$ on the phase space of the model.

\section{Branes in the symplectic language}

 Denote $SG$ a  "segment" group consisting od smooth maps from an interval $[0,\pi]$ into a (simple, connected, simply connected) compact Lie group $G$ and  consider  a space $SP$    of
 pairs $(g,J)$ where $g\in SG$ and $J\in Lie(SG)$. It makes thus sense to write down  a two-form  $\om_s$ and a vector field  $v_b$ on $SP$, given,  respectively, by the following formulas:

$$\om_{s}=d(J, dgg^{-1})_{b}+\jp k(dgg^{-1}\st \d_{\si}(dgg^{-1}))_{b}.\eqno(3)$$
$$v_b=\int^\pi_{0} \d_\si  J(\si){\delta\over \delta J(\si)} +{1\over k}\bigtriangledown^L_{J_L}-{1\over k}\btd^R_{J_R}.\eqno(4)$$
 Here  $dgg^{-1}$ is  a $Lie(SG)$-valued  right-invariant  Maurer-Cartan form on $SG$,
 $J_L$ and $J_R$ are defined as before and   $(.,.)_b$  stands for
$$(\chi_1,\chi_2)_{b}\equiv \bdr  \Tr (\chi_1(\si)\chi_2(\si)).$$ 
The "segment" formulae  (3),(4)  resemble  "loop"  formulae (1), (2)  but they do not define a dynamical system  since  the form $\om_s$ is not closed. Indeed, a simple book-keeping of boundary terms gives  
 $$d\om_s=-{k\over 6\pi} \Tr (dg(\pi)g(\pi)^{-1})^{ 3}+{k\over 6\pi}\Tr (dg(0)g(0)^{-1})^{ 3}\eqno(5)$$
 In order to modify $\om_s$ and render it closed,   we have to remind some standard concepts from the Lie group theory:
 \vskip1pc
 \noindent Let $\xi$ be an involutive automorphism of $G$, which respects also the Killing-Cartan form $\Tr(.,.)$ on $Lie(G)$. Consider then  two fixed elements $m_\pi,m_0\in G$   and their   corresponding  $\xi$-conjugacy classes $C^\xi_{m_\pi}, C^\xi_{m_0}$ in $G$. Recall that 
$$C^\xi_{m}=\{g\in G; \exists s\in G ,  g=\xi(s){m} s^{-1}\}.$$
\vskip1pc
\noindent To the element $m\in G$ and to the automorphism $\xi$, we can associate a two-form on $G$   given by
$$\ti \al^\xi _m=2\Tr ( \xi(s^{-1}ds) m s^{-1}ds m^{-1}),\eqno(6)$$
where we denoted the elements of $G$ by the symbol $s$.    We define also  a $\xi$-centralizer as
 Cent$^\xi(m)=
\{t\in G; \xi(t) m t^{-1}=m\}$. It is easy to check that the form
$\ti \al^\xi _{m}$ is  degenerated in the directions of  the  right action of  Cent$^\xi(m)$ on $G$ hence $\ti\al^\xi _{m}$ is a pull-back (under the map $s\to \xi(s)m s^{-1}$ ) of certain form $\al^\xi _{m}$ defined on the $\xi$-conjugacy class $C^\xi_{m}$  \cite{AS,Sta1, GTT}. 

Starting from the formula (6), it is easy to find out that 
$$d\al^\xi_{m}= {2\over 3}\Tr (dgg^{-1})^{ 3},\eqno(7)$$
where $g=\xi(s) m s^{-1}$.
\vskip1pc
\noindent The formulae (5) and (7)    motivate us to define a dynamical system $(M^\xi,\om^\xi_{m_\pi m_0},v_b)$ whose phase space is the submanifold
$M^\xi$ of  $SP$ such that 
$$ g(0)\in C^\xi_{m_{0}}, \quad  g(\pi)\in C^\xi _{m_{\pi}}, \eqno(8)$$ 
$$\xi(J_R(0)) =J_L(0), \quad \xi(J_R(\pi))=J_L(\pi), \eqno(9)$$
the  symplectic form  of which is
$$\om^\xi_{m_\pi m_0}=d(J, dgg^{-1})_{b}+\jp k(dgg^{-1}\st \d_{\si}dgg^{-1})_{b} +{k\over 4\pi}\al^\xi _{m_\pi}(g(\pi))-{k\over 4\pi} \al^\xi_{m_0}(g(0))$$
and the evolution vector field of which is 
$$v_b=\int^\pi_{0} \d_\si  J(\si){\delta\over \delta J(\si)} +{1\over k}\bigtriangledown^L_{J_L}
-{1\over k}\btd^R_{J_R}.$$
It is clear that the boundary conditions (8) (restricting the end-points of the string on the $\xi$-conjugacy classes)  make the form $\omm^\xi$ closed. The  glueing conditions (9) are dictated by the requirement:
that the evolution vector field $v_b$  must  respect the boundary conditions (8).   It turns out that 
$v_b$ is a hamiltonian vector field, i.e. $\iota_{v_b}\omm^\xi =dH_b$ with
 $$H_b=- {1\over 2\pi k}\int^\pi_0 \Tr(J_L^2+J_R^2) .$$
 This can be checked by using the following identity (cf. \cite{AMM}):
$$\iota_{v_b} \al^\xi_{m_0}= -{2\over k}\Tr(J_L(0),dg(0)g(0)^{-1}+\xi (g(0)^{-1}dg(0))) .$$

\section{Loop group parametrization}
Now we establish that the dynamical system $(M^\xi, \omm^\xi,v_b)$ introduced above  is nothing but the  boundary WZW model described in \cite{GTT}. For this, we first replace  the coordinates  $J(\si) , g(\si)$ on $M^\xi$  by a pair of  group valued variables  $g_R(\si), g_L(\si)$ as follows
$$J=-k\xi( g_R^{-1}\d_\si g_R), \quad g=\xi(g_R^{-1})g_L.$$ In the new parametrization, the symplectic form $\omm^\xi$ becomes
$$\omm^\xi=\jp k (dg_Lg_L^{-1}\st \d_\si (dg_Lg_L^{-1}))_b - \jp k (dg_Rg_R^{-1}\st \d_\si (dg_Rg_R^{-1}))_b     $$
$$-{k\over 2\pi}\Tr(\xi(dg_R g_R^{-1})\st dg_Lg_L^{-1})\hr +{k\over 4\pi}\al^\xi_\mu(\xi(g_R^{-1})g_L)\hr $$
 Note that the form $\omm^\xi$ is now    degenerated along the vector fields corresponding to the (gauge) transformations 
$(g_R(\si),g_L(\si))\to(\xi(A)g_R(\si),Ag_L(\si))$ with $A\in G$.  This degeneracy   expresses the ambiguity of the replacing $(J,g)$ by $(g_R,g_L)$. The boundary and the glueing conditions  now read
$$g_L(\pi)g_R^{-1}(\pi)\in  C^\xi_{m_{\pi}}, \quad g_L(0)g_R^{-1}(0)\in  C^\xi_{m_{0}},\eqno(10)$$
$$(g_R^{-1}\d_\si g_R)(\pi)=- (g_L^{-1}\d_\si g_L)(\pi),\quad (g_R^{-1}\d_\si g_R)(0)=-(g_L^{-1}\d_\si g_L)(0).\eqno(11)$$
\vskip1pc
\noindent  Set 
$$H_\pi \equiv g_L(\pi)g_R(\pi)^{-1}, \quad H_0\equiv g_L(0)g_R(0)^{-1}, 
 \quad G_0e^{2\pi\nu}G_0^{-1}=H_\pi H_0^{-1}$$
 where $\nu$ is in the Weyl alcove $A_+$.  We now introduce a $G$-valued field $h(\si)$, $\si\in [-\pi,\pi]$
 as follows
 $$h(\si)=g_L(\si)^{-1}G_0e^{\nu\si}, \quad {\rm for}\ \si\in [0,\pi],$$
 $$h(\si)=g_R(-\si)^{-1}H_0^{-1}G_0e^{\nu\si}, \quad {\rm for}\ \si\in [-\pi,0],$$
 In the new parametrization, the boundary   conditions (10) say that $h(\si)$ is a continuous loop,
 i.e. $h(-\pi)=h(\pi)$, and the glueing conditions (11) insure that $h(\si)$ is moreover smooth (in particular in $0$ and in $\pi$). The  form $\omm^\xi$ becomes
$$\omm^\xi=k(h^{-1}dh\st\d_\si(h^{-1}dh) )_\G+2kd(\nu,h^{-1}dh)_\G +$$
$$+ 2k\Tr(G_0^{-1}dG_0\wedge d\nu)+{k\over 2\pi}\Tr(H_\pi^{-1}dH_\pi \st H_0^{-1}dH_0)+$$
$$+ {k\over 4\pi}\al^\xi_{m_\pi}(H_\pi)- {k\over 4\pi}\al^\xi_{m_0}(H_0)-{k\over 2\pi}\Tr(G_0^{-1}dG_0e^{2\pi\nu}G_0^{-1}dG_0 e^{-2\pi\nu}).\eqno(12)$$
The degeneracy directions $(g_R(\si),g_L(\si))\to(\xi(A)g_R(\si),Ag_L(\si))$ of $\omm^\xi$  become 
now $(h(\si), \nu,H_\pi,H_0,G_0)\to (h(\si), \nu,AH_\pi \xi(A)^{-1}, AH_0\xi(A)^{-1}, AG_0)$.
The form $\omm^\xi$  given  by (12) coincides with the symplectic form of the boundary WZW model reported in the paper \cite{GTT}. The careful reader will notice, however, that the role of our boundary $\si=0$ is played by $\si=\pi$ in \cite{GTT}.

\section{Obstruction to duality}
 The loop group representation (12) shows  that  models with different $\xi$ may differ  just in the zero mode sector.  However, it is not easy to compare them directly since the
 ranges of the zero modes $(H_0,H_\pi, G_0,\nu)$ also differ for different $\xi$.  We shall therefore
 make once again a suitable change of parametrization  of the phase space of the $\xi$-boundary model. We first use a formula
$$\al^\xi_{\mu}(\xi(g)^{-1}hg)=$$ $$ \al_{\mu}^\xi(h) +2\Tr (\xi(dg g^{-1})\w
 hdgg^{-1} h^{-1}) -2\Tr(h^{-1}dh\w dgg^{-1} +dhh^{-1}\w \xi(dg g^{-1}))$$
and then rewrite the form $\omm^\xi$ as follows:
$$\omm^\xi=\jp k(dg_Lg_L^{-1}\st\d_\si(dg_Lg_L^{-1}))_b-\jp k(dg_Rg_R^{-1}\st \d_\si(dg_Rg_R^{-1})_b$$
$$+{k\over 2\pi}\Tr\biggl(g_\pi^{-1}dg_\pi\w (\xi( dg_L(\pi)g_L(\pi)^{-1}) - dg_R(\pi)g_R(\pi)^{-1}) \biggr)$$ $$-{k\over 2\pi}\Tr\biggl(g_0^{-1}dg_0\w (\xi( dg_L(0)g_L(0)^{-1}) - dg_R(0)g_R(0)^{-1})\biggr) .\eqno(13)$$
Here  the variables $g_0$, $g_\pi$ are defined by 
$$g_L(\pi)g_R(\pi)^{-1}=\xi(g_\pi)^{-1}m_\pi g_\pi, \quad g_L(0)g_R(0)^{-1}=\xi(g_0)^{-1}m_0g_0.\eqno(14)$$
 Set 
$$\ti g_{R}(\si)=p^{-1}e^{\tau\si}pg_0g_{R}(\si)$$
 $$\ti g_{L}(\si)=\xi(p^{-1}e^{\tau\si} pg_0)g_{L}(\si)$$
  where   $ p\in G$, $\tau\in A_+$  are defined as 
 $$  p^{-1}e^{\tau\pi}p=g_\pi g_0^{-1}.\eqno(15)$$
The  symplectic  form $\omm^\xi$ (xy) then becomes 
$$\omm=\jp k  (d\ti g_L\ti g_L^{-1}\st \d_\si (d\ti g_L\ti g_L^{-1}))_b -
 \jp k (d\ti g_R\ti g_R^{-1}\st \d_\si (d\ti g_R\ti g_R^{-1}))_b    + $$
 $$+kd(\xi(\chi), d\ti g_L\ti g_L^{-1})_b  -kd(\chi, d\ti g_R\ti g_R^{-1})_b,$$
 where $$\chi = p^{-1}\tau p.\eqno(16)$$
 The Hamiltonian now reads
 $$H^\xi_b=-{k\over 2\pi}\int^\pi_0d\si \Tr\biggl((\d_\si \ti g_L\ti g_L^{-1}-\xi(\chi))^2  +( \d_\si \ti g_R\ti g_R^{-1}-\chi)^2 \biggr).$$
 The boundary and glueing conditions (10, 11) become, respectively, 
 $$\ti g_L(0)=m_0  \ti g_R(0), \quad \ti g_L(\pi)= m_\pi \ti g_R(\pi), $$
 $$ \biggl( \ti g_L^{-1}\d_\si \ti g_L -\ti g_L^{-1}\xi(\chi) \ti g_L\biggr)_{\si=0,\pi}= 
 -\biggl(\ti g_R^{-1}\d_\si \ti g_R -
 \ti g_R^{-1}\chi\ti g_R\biggr)_{\si=0,\pi}.$$
 In the last parametrization, we observe that the range of the new  variables $(\ti g_L,\ti g_R,\chi)$ is the same for all automorphisms $\xi$, in particular $e^{\pi\chi}$ sweeps the whole group $G$. This means that 
 the observable $\chi$ is the obstruction for T-duality since,  in general,  we have $\xi_1(\chi)\neq \xi_2(\chi)$ 
 for $\xi_1\neq\xi_2$. Indeed, the  T-duality means the dynamical equivalence of two models,  therefore it cannot be   established unless we impose  the constraint  $\xi_1(\chi)=\xi_2(\chi)$ for {\it both} models.  It seems that we cannot do better and the T-duality takes place only between the pair of  the {\it  constrained} models. Generically,  this conclusion is correct but there are particular brane
 geometries  for  which  the constraint  $\xi_1(\chi)= \xi_2(\chi)$ is satisfied dynamically and need not be imposed by hand.   We shall explain in the next section how this happens and we shall interpret the phenomenon as the nested duality.
 
 \section{Non-regular branes}
 
 It is important to notice that, so far,  we have been somewhat abusing the terminology by calling "symplectic"    also those 2-forms which might have been only presymplectic. 
Here by adjective "presymplectic" we mean a 2-form which is closed but need not be non-degenerate.
The possible degeneracy  of a  closed 2-form $\omega$ does not mean that we do not deal with a honest dynamical
system  but it rather indicates that we study the system with a gauge symmetry. Infinitesimally,
the gauge symmetry is generated by vector fields $w$ on the phase space that annihilate the symplectic
form, i.e. such that $\iota_w\omega=0$. It is the  closedness of the form $\omega$ then insures that 
the Lie bracket $\{w_1,w_2\}$ of two annihilating vector fields is again an annihilating vector
field.  

In our context, the degeneracy gauge group $K^\xi_{\mu_\pi\mu_0}$ {\it depends} on the brane geometry and also  on  the model attributed to $\xi$ and it is given as the
direct product of the  twisted centralizers, i.e.
$K^\xi_{m_\pi m_0}=$ Cent$^\xi_{m_\pi}\times$Cent$^\xi_{m_0}$. This fact can be easily seen from (14) since it expresses the ambiguity of the definition  of the variables $g_\pi,g_0$. Indeed, the pair
$(t^\xi_\pi g_\pi,t^\xi_0g_0)$ is equivalent to $(g_0,g_\pi)$ for $t^\xi_{\pi,0}\in$Cent$^\xi_{m_{\pi,0}}$ and the reader may check by direct calculation the corresponding degeneracy of the symplectic form (13). Now the variable $\chi$ transforms nontrivially under the gauge transformations:
$$e^{\pi\chi}\to t^\xi_\pi e^{\pi\chi} (t^\xi_0)^{-1},\eqno(17)$$
as it can be seen from Eqs. (15) and (16).

 We  ask the following question:
If we work with the $\xi_1$-model, can we achieve  the
equality  $\xi_1(\chi)=\xi_2(\chi)$ by a $K^{\xi_1}$-gauge transformation?  If yes, then we see immediately that the $\xi_1$-boundary WZW model is 
the dynamical subsystem of the $\xi_2$-boundary WZW model selected by the constraint $\xi_2(\chi)=\xi_1(\chi)$
imposed on the {\it latter}.   Similarly,  if we work with the $\xi_2$-model and we achieve
the equality   $\xi_1(\chi)=\xi_2(\chi)$ by $K^{\xi_2}$-gauge transformation we conclude that the
$\xi_2$-model is the subsystem of the $\xi_1$-model. Of course, the ideal situation would
be, if for some choice $m_\pi,m_0$ the equality $\xi_1(\chi)=\xi_2(\chi)$  could be reached by both $K^{\xi_1}$ and $K^{\xi_2}$
gauge transformations. This would mean that the  $\xi_1$-model and $\xi_2$ model would be strictly
dynamically equivalent  or, in other words, T-dual to each other. Unfortunately, we did not find such 
a configuration $m_\pi,m_0$ and we even conjecture that it does not exist. Nevertheless, the
situation when one model is the submodel of the other does happen for the so called non-regular branes.  In what follows, we shall illustrate the ideas and concepts of this section on the case
of the group $G=SU(3)$. The first automorphism $\xi_1=id$ will be just trivial identity map and the
second $\xi_2=cc$ will be given by the standard complex conjugation.

We start with the twisted  case (i.e. $\xi_2 =cc$) and we wish to find a gauge transformation which would
bring any value of the variable $\chi$  onto the slice $\chi^{cc}=\chi$.  This means that the gauge
group $K^{cc}_{m_\pi m_0}$ has to be at least 5-dimensional. Indeed, the subgroup of 
$cc$-invariant elements (i.e. real matrices) in $SU(3)$ is three dimensional while the dimension
of $SU(3)$ itself is eight.  The possible twisted centralizers Cent$^{cc}(m)$ have been classified by Stanciu in \cite{Sta2}. They are mostly one-dimensional but in two cases they have dimension
three. In particular, it is the $cc$-conjugacy class of the unit element for which  Cent$^{cc}(e)=SO(3)$
and the $cc$-conjugacy class of the element  $f\in SU(3)$
$$f=\left(\matrix{0&1&0\cr -1&0&0\cr 0&0&1}\right)$$
for which Cent$^{cc}(f)=SU(2)$.  Thus the counting of dimensions says that it is necessary for both string end-points  to live either on $C^{cc}(e)$ or on $C^{cc}(f)$ if we want to get to the slice $\chi^{cc}=\chi$
by the gauge transformation. Unfortunately, this condition  is not sufficient.  Indeed, the centralizer of the unit element
does not take $\chi$ away from the slice (since it is real) therefore it cannot bring $\chi$ on the slice either. 
It remains only the case when both string end-points live on  $C^{cc}(f)$. In this case Cent$^{cc}(f)$ is embedded in $SU(3)$ in the following way:
$$\left(\matrix{a&b&0\cr -\bar b&\bar a&0\cr 0&0&1}\right), \quad a\bar a +b\bar b=1.\eqno(18)$$
It is clear that a  one-parameter group of real matrices
$$\left(\matrix{\cos{\theta}&\sin{\theta}&0\cr -\sin{\theta}&\cos{\theta}&0\cr 0&0&1}\right)$$
is contained in Cent$^{cc}(f)$ and, at the same time, it cannot help to bring  $\chi$ on the slice. Thus 
we loose one parameter  at each end-point and the four
remaining parameters from two copies of Cent$^{cc}(f)$ cannot bring the 8-dimensional object $\chi$
onto the 3-dimensional slice  $\chi^{cc}=\chi$. We  conclude that the twisted $cc$-boundary WZW
model is never a subsystem of the ordinary $id$-boundary WZW model, whatever is the  choice 
of the  $cc$-conjugacy classes.

We now show, on the contrary, that  the ordinary $id$-boundary WZW model can be the subsystem
of the $cc$-boundary WZW model or, in other words, the nested duality can take place.  Indeed,   the slice  $\chi^{cc}=\chi$ can be achieved by the gauge
transformation  from $K^{id}_{m_\pi m_0}$  for several conjugacy classes $C^{id}_{m_\pi}$ and
$C^{id}_{m_0}$. In particular, if  we  consider a point-like brane $C^{id}_c$ where the  element $c$ is from the $SU(3)$ center, we have Cent$^{id}(c)=SU(3)$ and the formula
(xy) says that  the gauge transformation can bring any $\chi$  on the slice  $\chi^{cc}=\chi$. 
There is another example, which si perhaps more interesting. Consider  a fixed  real number $\phi$
and the following element of $SU(3)$. 
$$m_\phi=\left(\matrix{e^{{2\pi i\over 3}\phi} &0&0\cr 0&e^{{2\pi i\over 3}\phi} &0\cr 
0&0&e^{-{4\pi i\over 3}\phi}}\right).\eqno(19)$$
The conjugacy class $C^{id}_{m_\phi}$ is 4-dimensional and it  is  often refered to as being non-regular because the centralizer
Cent$^{id}(m_\phi) $ is non-Abelian.  Indeed, Cent$^{id}(m_\phi) =SU(2)\times U(1)$ where $SU(2)$
is embedded in $SU(3)$ as in (18) and $U(1)$ is embedded as (19) with $\phi$ varying.    If both 
string end-points live on $C^{id}_{m_\phi}$ then the gauge group $K^{id}_{m_\phi m_\phi}=
SU(2)\times SU(2)\times U(1)\times U(1)$ is 8-dimensional and the naive counting of dimensions
gives a hope to bring $\chi$ on the slice $\chi^{cc}=\chi$. To see that this indeed happens we use the
following decomposition (cf. \cite{RSG}) of the general $SU(3)$ element:
$$e^{\chi}= h_L\left(\matrix{1&0&0 \cr 0&e^{i\al}&0\cr  0&0&e^{-i\al}}\right)\left(\matrix{1&0&0 \cr 0&\cos{\beta}&- \sin{\beta}\cr  0&\sin{\beta}&\cos{\beta}}\right)\left(\matrix{1&0&0 \cr 0&e^{i\al}&0\cr  0&0&e^{-i\al}}\right)h_R,$$
where $h_{L,R}$ are of the form (18).
Since the matrix in the middle of the decomposition is real and  all other matrices are elements of Cent$^{id}(m_\phi)$ we see that every $\chi$ can be brought onto the slice $\chi^{cc}=\chi$ by the
gauge transformation (17). 

We finish with two  comments on   quantization: 1) First of all, among the branes  leading to the nested duality there are such that satisfy the quantization condition of the single-valuedness of the path integral
(cf. \cite{KS,AS}). This is certainly true for the case where one end-point of the open string sticks on
the element $c$ of the group center but also for the case when both end-points are attached to 
$C^{id}_{m_\phi}$. Indeed, in the latter case, it is sufficient to look at Figure 4 of Ref. \cite{Sta2} where
it is described the quantum moduli space of  $id$-branes for several lowest levels $k$ (they coincide
with the affine dominant weghts and are depicted by small black disks). With varying $\phi$, our conjugacy class  $C^{id}_{m_\phi}$ sweeps  the boundary  of  the triangular  Weyl alcove, hence it intersects the quantum moduli space.  2) Another quantum issue concerns the effective
field theory of the open string massless modes.  In particular, it would be interesting to relate
the results of \cite{ARS} and of \cite{AFQS} via the nested T-duality.


\begin{thebibliography}{19}
\bibitem{KS}{C. Klim\v c\'\i k and P. \v Severa, {\it Open strings and D-branes in WZNW models}, Nucl. Phys. {\bf B488} (1997) 653-676 [hep-th/9609112]}
\bibitem{KO}{E. Kiritsis and N. Obers, {\it A New Duality Symmetry in String Theory}, Phys.Lett. {\bf  B334} (1994) 67-71 [hep-th/9406082] }
\bibitem{KS1}{C. Klim\v c\'\i k and P. \v Severa, {\it Non-Abelian momentum-winding exchange}, Phys. Lett.  {\bf B383} (1996) 281-286 [hep-th/9605212]}
\bibitem{KP}{C. Klim\v c\'\i k and S. Parkhomenko, {\it The Poisson-Lie T-duality and zero modes}, 
 Theor. Math. Phys.{\bf 139} (2004) 834-845 [hep-th/0010084]}
\bibitem{M}{J. Balog, L. Feh\'er and L. Palla, {\it Chiral Extensions of the WZNW Phase Space, Poisson-Lie Symmetries and  Groupoids}, Nucl.Phys. {\bf B568} (2000) 503-542, [hep-th/9910046]}
\bibitem {K}  {C. Klim\v c\'\i k, {\it Quasitriangular 
WZW model},   Rev. Math. Phys. {\bf 16}  (2004) 679 - 808,  [hep-th/0103118]}
 \bibitem{G}{ÊK. Gaw\c edzki, {\it  Classical origin of quantum group symmetries in Wess-Zumino-Witten conformal field theory},  Comm. Math. Phys. {\bf 139}  (1991)  201--213}
\bibitem{GTT} {K. Gaw\c edzki, I. Todorov and  P. Tran-Ngoc-Bich, {\it Canonical quantization of the boundary Wess-Zumino-Witten model} , Commun.Math.Phys. {\bf 248} (2004) 217-254 
 [hep-th/0101170]}
 \bibitem{Gaw}{ K. Gaw\c edzki, {\it  Boundary WZW, G/H, G/G and CS theories},  Annales Henri Poincare {\bf 3}  (2002) 847-881 [hep-th/0108044]}
Ê\bibitem{AS}{A. Alekseev and V. Schomerus, {\it D-branes in the WZW model}, Phys.Rev. {\bf D60} (1999) 061901 [hep-th/9812193]}
\bibitem{Sta1}{S. Stanciu, {\it A note on D-branes in group manifolds: flux quantisation and D0-charge}, JHEP {\bf 10}  (2000) 015. [hep-th/0006145]}
\bibitem{AMM}{A. Yu.  Alekseev, A. Z. Malkin, E. Meinrenken, {\it  Lie Group Valued Moment Maps},
J. Differential Geom. {\bf 48} (1998) 445--495 [dg-ga/9707021]}
\bibitem{Sta2}{S. Stanciu, {\it An illustrated guide to D-branes in SU(3)}, [hep-th/0111221]}
\bibitem{RSG} {D. J. Rowe, B. C. Sanders and   H. de Guise, {\it Representations of the Weyl group and Wigner functions for SU(3)}, Journal Math. Phys. {\bf 40} (1999) 3604-3615}
\bibitem{ARS} A. Yu. Alekseev, A. Recknagel and V. Schomerus, {\it  Brane dynamics in background fluxes and non-commutative geometry}, JHEP {\bf 05} (2000) 010 [hep-th/0003187].  
\bibitem{AFQS}{ A. Yu. Alekseev, S. Fredenhagen, T. Quella and V.Schomerus,
{\it Non-commutative gauge theory of twisted D-branes},  Nucl.Phys. {\bf B646}  (2002) 127-157,
[hep-th/0205123]}

 \end{thebibliography}
\end{document}